\documentclass[aps,prl,twocolumn,superscriptaddress]{revtex4}
\usepackage{natbib}
\usepackage{amsmath}
\usepackage{graphicx}
\usepackage{dcolumn}
\usepackage{tabularx}
\usepackage{keyval}
\usepackage{multirow}

\usepackage{booktabs}
\usepackage{threeparttable}

\begin{document}
\title{Hydrostatic pressure induced three-dimensional Dirac semimetal in black phosphorus}
\author{Peng-Lai Gong}
\affiliation{Key Laboratory of Materials Physics, Institute of Solid State Physics,
Chinese Academy of Sciences, P. O. Box 1129, Hefei, 230031, China}
\author{Da-Yong Liu}
\email{dyliu@theory.issp.ac.cn}
\affiliation{Key Laboratory of Materials Physics, Institute of Solid State Physics,
Chinese Academy of Sciences, P. O. Box 1129, Hefei, 230031, China}
\author{Kai-Shuai Yang}
\affiliation{Key Laboratory of Materials Physics, Institute of Solid State Physics,
Chinese Academy of Sciences, P. O. Box 1129, Hefei, 230031, China}
\author{Zi-Ji Xiang}
\affiliation{Department of Physics, University of Science and Technology of China,
Hefei, 230026, China}
\author{Xian-Hui Chen}
\affiliation{Department of Physics, University of Science and Technology of China,
Hefei, 230026, China}
\author{Zhi Zeng}
\affiliation{Key Laboratory of Materials Physics, Institute of Solid State Physics,
Chinese Academy of Sciences, P. O. Box 1129, Hefei, 230031, China}
\affiliation{University of Science and Technology of China, Hefei, 230026, China}
\author{Shun-Qing Shen}
\affiliation{Department of Physics, The University of Hong Kong, Pokfulam Road,
Hong Kong, China}
\author{Liang-Jian Zou}
\email{zou@theory.issp.ac.cn}
\affiliation{Key Laboratory of Materials Physics, Institute of Solid State Physics,
Chinese Academy of Sciences, P. O. Box 1129, Hefei, 230031, China}
\affiliation{University of Science and Technology of China, Hefei, 230026, China}
\date{\today}

\begin{abstract}
We present the first-principles studies on the hydrostatic pressure effect of the electronic
properties of black phosphorus. We show that the energy bands crossover
around the critical pressure $P_{c}$=1.23 GPa; with increasing pressure, the band reversal occurs
at the $Z$ point and evolves into 4 twofold-degenerate Dirac cones around the $Z$ point,
suggesting that pressured black phosphorus is a 3D Dirac semimetal.
With further increasing pressure the Dirac cones in the $\Gamma$-$Z$ line move
toward the $\Gamma$ point and evolve into 2 hole-type Fermi pockets, and those in the
$Z$-$M$ lines move toward the $M$ point and evolve into 2 tiny electron-type Fermi pockets, and
a band above the $Z$-$M$ line sinks below E$_{F}$ and contributes 4 electron-type pockets.
A clear {\it Lifshitz} transition occurs at $P_{c}$ from semiconductor to 3D Dirac
semimetal. Such a 3D Dirac semimetal is protected by the nonsymmorphic space symmetry of bulk
black phosphorus.
These suggest the bright perspective of black phosphorus for optoelectronic and electronic devices
due to its easy modulation by pressure.

\end{abstract}


\maketitle

\section{I. Motivations}
The {\it van der Waals} force between layers makes layered compounds  graphite, MoS$_{2}$, WTe$_{2}$
and TaS$_{2}$, {\it etc.},be easily exfoliated to monolayers as
two-dimensional materials, such as graphite to graphene and MoS$_{2}$ to its monolayer.
As a possible candidate of optoelectronic and electronic material, the energy gap of black
phosphorus varies from 0.3 eV in bulk to about 2 eV in monolayer, filling
the optical interval between small energy gap 0-0.3 eV in graphene and large energy gap of 1-2 eV in
semiconductive dichalcogenides \cite{Xia2014}.
Recently, black phosphorus in
bulk, multilayers and monolayers have received considerable interests
\cite{low2014plasmons,rodin2014strain,Vy2014Layer}.
Zhang {\it et al.} demonstrated that multilayer black phosphorus exhibits as high
as 10$^{5}$ drain current modulation and 10$^{3}$ $cm^{2}V^{-1}s^{-1}$ charge
mobility \cite{li2014black}, showing that black phosphorus thin film might be a good
potential candidate for field effect transistor \cite{xia2014rediscovering}.
More recently it is shown that a moderate hydrostatic pressure could not only drive
semiconductive black phosphorus to metallic, but also tune multiple Fermi surfaces and Lifshitz
point in the magnetotransport and
Shubnikov-De Haas oscillation measurements \cite{xiang2015pressure}.
These properties make black phosphorus a new exciting field
both in material sciences and condensed matter physics \cite{YePD2015}.
%
%

Hydrostatic pressure could easily change the crystal structure of black phosphorus and
modify its electronic properties \cite{cartz1979effect,rodin2014strain}.
As the most stable allotrope of phosphorus, black phosphorus exhibits
three different phases under moderate high hydrostatic pressures: the orthorhombic phase
with wrinkled hexagons, the graphene-like rhombohedral phase with hexagonal
lattice for the pressure $P$$>$4.5 $GPa$ \cite{okajima1984electrical},
and the simple cubic phase
for $P$$>$10.3 $GPa$ \cite{okajima1984electrical}.
Under high pressure the P-P bond lengths and bond angle of black phosphorus display
strong pressure dependence, and exhibit highly anisotropy in compressibility \cite{cartz1979effect}.
Though a strain may crucially modify the energy gap of monolayer black phosphorus
\cite{rodin2014strain}, it is completely different from the hydrostatic pressure effect on bulk black phosphorus,
since the {\it van der Waals} force leads to considerable interlayer coupling \cite{Qiao2014High}.
One anticipates that a high
hydrostatic pressure significantly compresses the interlayer distance of P atoms, however, it remains a
puzzle and unclear scenario: how do the electronic structures evolve with increasing hydrostatic
pressures? Especially, how does the Lifshitz point occur under pressure?

On the other hand, searching for 3D Dirac semimetals
has attracted great attention in recent years \cite{wang2012dirac,gibson20143d}.
Recently Kim {\it et al}. claimed \cite{Kim2015} that they observed the Dirac semimetal state by depositing K
ions on the surface of black phosphorus, however, their Dirac cones near the $\Gamma$ point actually
always sink below E$_F$.
Cava {\it et al.} \cite{gibson20143d} pointed out that a crystal with glide planes and scream axes might form a
3D Dirac semimetal when the electronic structures of the crystal possess a twofold-degenerate
bands, since the nonsymmorphic symmetry in the crystal lattice leads to the band inversion at a point in
the Brillouin zone \cite{Zak1999,Zak2002Topologically}.
The space group of bulk black phosphorus has a glide plane,
thus it is a potential candidate of a 3D Dirac semimetal.
To elucidate electronic and optical properties of black
phosphorus under hydrostatic pressure, in this Letter, using the first-principles
electronic structures calculations and analytic studies,
we for the first time present the evolutions of the band structures and
Fermi surfaces of black phosphorus under hydrostatic pressure, and clearly demonstrate the $Lifshitz$ transition
from a semiconductor to a Dirac semimetal.
We argue that the non-symmorphic symmetry in crystal structure of black phosphorus protects it as a
3D Dirac semimetal.

\section{II. COMPUTATIONAL STRUCTURES AND PHONON SPECTRA}

\subsection{II.1 Computational details}

To explore the evolutions of crystal structures, electronic structures and physical properties with
hydrostatic pressures, we first achieve the stable crystal structures of black
phosphorus under various pressures within the framework of density functional theory (DFT)
using the projector augmented wave (PAW)
method \cite{paw1,Kresse1999From} as implemented in the Vienna Ab initio Simulation Package
(VASP) code \cite{Kresse1996Efficient}.  These structures are optimized by using
the {\it optB88-vdW} method \cite{optb88-1,optb88-2}, so as to take into account the interlayer
{\it van der Waals} force in bulk black phosphorus.

\begin{figure}[!b]
\noindent \begin{centering}
\includegraphics[width=7.0cm]{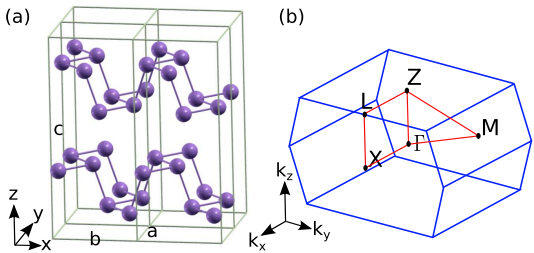}
\par\end{centering}
\caption{ Crystal structure of bulk black phosphorus (a) and the first
 Brillouin zone and some high symmetric points of bulk black phosphorus (b). }
\label{Fig1-crystal}
\end{figure}

The experimental lattice parameters of bulk black phosphorus are taken as the initial structure.
In our work, the directions of lattice constants $a$ and $b$ are set to be along  the zigzag
(the $y$ axis) and the armchair ( the $x$ axis)  directions, respectively, in the crystal structure
of bulk black phosphorus;
and $c$ is along the interlayer direction (the $z$ axis), as displayed in Fig. \ref{Fig1-crystal}.
Both the shape and volume of each supercell have been relaxed fully at a target pressure
and all atoms in the supercell were allowed to move until the residual force per atom was
drop below 0.001 eV/$\AA$. The energy cutoff for the plane-wave basis was set to
500 eV for all calculations. A k-mesh of 10$\times$ 4 $\times$ 8 in band structure calculation was
adopted to sample the first Brillouin zone of the primitive unit cell of bulk black phosphorus.
Uniform k-mesh of 30$\times$30$\times$30 was adopted to obtain Fermi surfaces when black
phosphorus changes into semimetal.
Phonon calculations were carried out using the finite displacement method and PHONONPY
package \cite{phononpy1,phononpy2}.  A 3$\times$3$\times$4 supercell  and a
5$\times$5$\times$5 k-mesh were applied to calculate the force constants.

\subsection{II.2 Crystal structure evolution and phonon spectra}

\begin{figure}[!b]
\noindent \begin{centering}
\includegraphics[width=4.0cm]{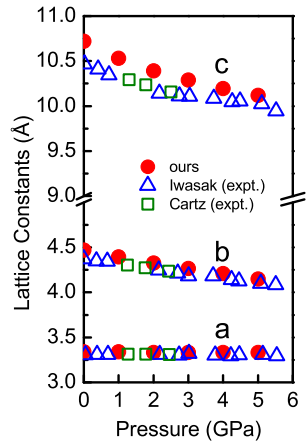}
\par\end{centering}
\caption{Lattice constants of bulk black phosphorus under various pressures. The experimental
data  from Iwasaki \textit{et al}. [24]  and Cartz  \textit{et  al}. [25] are  also  shown  as  open triangles
and squares  for comparison.}
\label{Fig2-constants}
\end{figure}

We find that the calculated lattice constants of bulk black phosphorus in the orthorhombic phase
under various pressures shown in Fig. \ref{Fig2-constants} are in agreement with the experimental
results \cite{Iwasaki1983,Cartz1979} very well. In detail, the calculated lattice constants $a$ and
$b$ are very close to the experimental results \cite{Iwasaki1983,Cartz1979}, though the lattice
constant $c$ is slightly larger than the measured within 0.2\%, since the GGA functionals with the
$vdW$-correction overestimates the layer distance $c$, implying that our theoretical structural
parameters are highly reliable.
Furthermore, the phonon spectra for our theoretical structures also demonstrate that the pressured
structures are stable. Our numerical results have shown that all of our theoretical structures have
not any imaginary frequency, as shown in  Fig. \ref{FigA1-phonon} in the APPENDIX.

\section{III. Electronic structures evolution under pressure}
Electronic band structures were obtained by the VASP method based on the different
pressurized structures.  All of the calculations are cross-checked by using the Quantum Espresso
\cite{Giannozzi2009QUANTUM} and WIEN2K \cite{blaha2001augmented} codes, respectively,
and the results are consistent with each other. Within the hybrid functional of Heyd,
Scuseria, and Ernzerhof (HSE) framework \cite{Heyd2003Hybrid,Heyd2006Hybrid}, we obtain an energy gap of 0.34 eV at
ambient condition, in agreement with the experimental and recent theoretical data \cite{li2014black,Qiao2014High}.
%
%

%

To save the computation resource and accord with the experiments \cite{xiang2015pressure},
throughout this paper we
present the numerical results within the modified Becke-Johnson (mBJ) exchange-correlation
potential \cite{mbj1,mbj2}  for the semiconducting phase, and the
PBE exchange-correlation potential \cite{pbe} for the metallic phase.
It is now recognized that similar to the HSE or GW methods, the mBJ method could well describe
a wide range of materials properties, including the band gap of semiconductors, while it sometimes
make false predictions of fundamental properties of metals \cite{ZhangCB2015}.
During the application of hydrostatic pressure, the black phosphorus undergoes from semiconductive
phase to semimetallic one. It is desirable and self-consistent to adopt the same first-principles method
with the same exchange-correlation potential to calculate the electronic properties of black phosphorus
over wide pressure range. However, due to the limit of each method, we have to adopt different
exchange-correlation potential to calculate the energy band of the semiconductor and semimetal.

\begin{figure}[!tb]
\noindent \begin{centering}
\includegraphics[width=6.0cm]{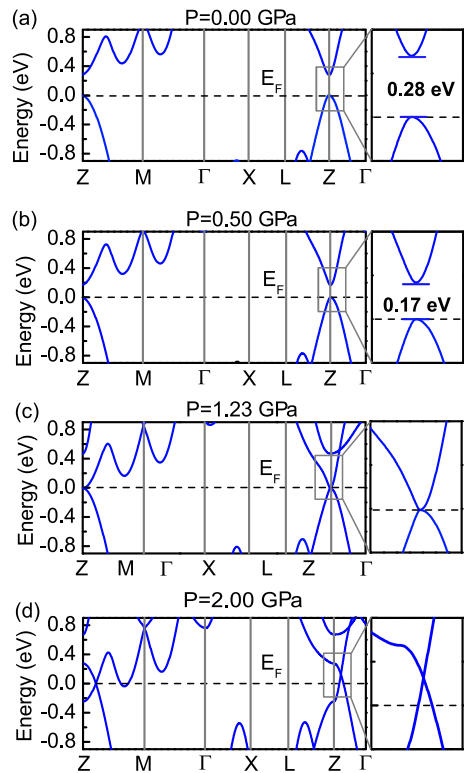}
\par\end{centering}
\caption{(Color Online) Evolution of energy  band structures of bulk black phosphorus with
increasing pressures from 0 GPa (a), 0.5 GPa (b), 1.23 GPa, the critical pressure for
band crossover (c),  2.0 GPa (d). The first three figures and the last one are the mBJ and PBE results, respectively.}
\label{Fig4-bands}
\end{figure}

The pressure dependence of band structures for the semiconductive (mBJ results) and for the
semimetallic (PBE result)  phases of bulk black phosphorus is shown
in Fig. \ref{Fig4-bands}. It shows that the system opens a gap about 0.28 eV
at 0 GPa. With increasing the pressure to 0.5 and 1.0 GPa, up to about $P_{c}$=1.23 GPa,
the band gap gradually closes and the two bands below and above E$_F$ touch at the Z-point.
Thus the critical pressure for the semiconductor-semimetal transition is about $P_{c}$=1.23 GPa
in the mBJ functional, which is comparable with the value
observed in the recent experiment \cite{xiang2015pressure}. Since the energy bands begin to
anti-inverse at $P_{c}$, the black phosphorus becomes a semimetal with a degenerate Dirac point
at the $Z$ point, as seen the band structures in Fig. \ref{Fig4-bands}(c).

\begin{figure}[!htbp]
\noindent \begin{centering}
\includegraphics[width=6cm]{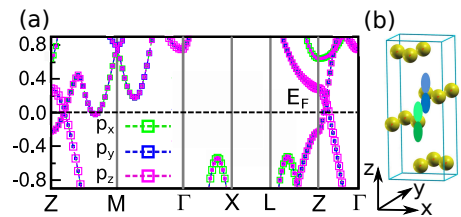}
\par \end{centering}
\caption{(Color online) Bands structure with orbital projection of bulk black phosphorus at 2.0 GPa (a) and schematic diagram
of $3p_z$ orbital distribution contributed to the Dirac cones (b).}
\label{FigA2-pdos}
\end{figure}

When the hydrostatic pressure increases up to 1.5  GPa (not shown), the energy bands already cross with each
other  around the Fermi energy E$_{F}$, the anti-inversed energy bands develop
the linear dispersions and Dirac points
in the $Z$-$M$ and $Z$-$\Gamma$ directions, respectively.
We attribute to that $p_{z}$ orbitals along the $c$ axis gradually overlap with
increasing the pressure, as shown in Fig. \ref{FigA2-pdos}, majorly contributing to the band anti-inversion.
At a pressure of 2.0 GPa shown in Fig. \ref{Fig4-bands}(d), the Dirac cones in the $Z$-$\Gamma$ direction
shift toward the $\Gamma$ point and develops two hole Fermi pockets. On the other hand,
the Dirac cones in the $Z$-$M$ line shift toward the $M$ point
and above E$_F$ in energy, contributing
2 tiny electron pockets; meanwhile a band above the $Z$-$M$ line sinks below E$_F$ and develops 4 electron
Fermi pockets. At pressures up to 2.5 $\sim$ 4 GPa we studied, the volumes of the hole
Fermi pocket gradually expands, and an electron Fermi pocket in the $Z$-$M$ line
becomes considerably large, in agreement with recent experimental observation \cite{xiang2015pressure}.
Thus it is naturally expected that the hole and electron
carriers should coexist under high pressure until 4 GPa, since high hydrostatic
pressure does not break the charge balance.

As well known, the wide bandwidth 2$p_{z}$ orbitals of carbon atoms contribute the six Dirac cones of
graphene. One may wonder which orbital or which a few orbitals contribute the Dirac cones when the hydrostatic
pressure $P$$>$$P_{c}$. After projecting the weight of orbitals, we could find that the $3p_z$ orbitals contribute
the dominant weight of the Dirac cones. As seen in Fig. \ref{FigA2-pdos}, under the high hydrostatic pressure,
the $3p_{z}$ orbitals overlap with each other, and lead to the linearized spectrum.

\section{IV. Dirac cones and 3D Dirac semimetal:}
\begin{figure}[!b]
\noindent \begin{centering}
\includegraphics[width=6.0cm]{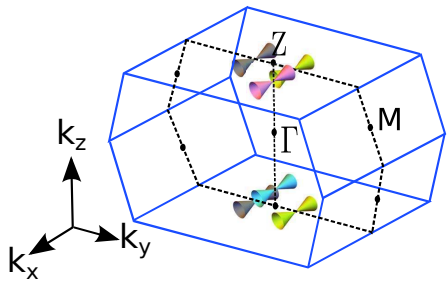}
\par\end{centering}
\caption{(Color Online) Sketched distribution of 4 Dirac cones in the first Brillouin zone of
bulk black phosphorus under the pressure of 2.0 GPa. The two Dirac cones lie on the
upper and lower surfaces of the Brillouin zone are identical.}
\label{Fig6-dirac}
\end{figure}

From the band structures one finds that when the hydrostatic pressure increases to the critical value of
$P_{c}$=1.23 GPa, the valence band and the conduction band touch together at the high-symmetric $Z$
point of the Brillouin zone and gradually intersect with each other. The energy
band structures of the system exhibit anti-inversion under further pressure. As described above,
when the pressure becomes larger than the critical point, $P$ $>$ $P_c$, the anti-inverted points at the
high-symmetric $Z$ point evolve into Dirac points around the $Z$ point, and this fourfold-degenerate
point gradually splits into four Dirac cones with the increase of pressure, as sketched in Fig. \ref{Fig6-dirac}
and confirmed by Fermi surfaces evolution under pressures shown later. Taking into account the spin
degree of freedom, each Dirac cone remains twofold degenerate.
Note that these Dirac cones are twofold-degenerate once taking into account the spin freedom of degree,
thus the influence of magnetic field deserves for further investigation.

\begin{figure}[!tb]
\noindent \begin{centering}
\includegraphics[width=6.0cm]{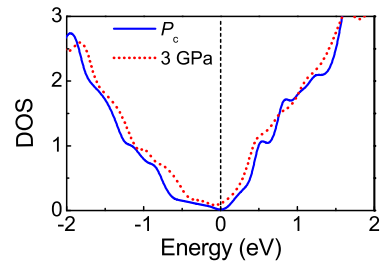}
\par\end{centering}
\caption{(Color Online) Density of states (DOS) of bulk black phosphorus under
the pressure of $P_{c}$=1.23 GPa (blue solid line)  and 3.00 GPa (red dashed line).}
\label{Fig7-dos}
\end{figure}

The 3D Dirac semimetal can further be verified from the
density of states (DOS) of black phosphorus near the critical pressure  $P_c$ shown in
Fig. \ref{Fig7-dos}. One observes that the DOS vanishes near E$_F$,
suggesting well-defined semimetallic nature. These properties strongly indicate that bulk black
phosphorus under intermediate pressure is the first realistic example of
pressure-modified 3D Dirac semimetal. At $P$=3.0 GPa shown in Fig. \ref{Fig7-dos},
finite DOS appears near E$_F$  due to the upward and downward of bands, implying
that the system becomes metallic. However, the DOS
near E$_F$ does not increase considerably.

The appearance of the band inversion, Dirac points and Dirac cones in black phosphorus
are particularly interesting, since in this calculation, neither considerable spin-orbit
coupling is taken into account, nor a graphene-like honeycomb structure is realized
in black phosphorus
, though the crystal structure of black phosphorus changes significantly.
This raises a question: what is the physical origin of the formation of the  Dirac semimetal?
Considering the space-group symmetry operators of the black
phosphorus, we attribute the origin of Dirac points to
the symmetry of the nonsymmorphical space group in black phosphorus.
In the hydrostatic pressure range
we studied, the space group of the crystal structure of black phosphorus is always
$Cmca$, which possesses a nonsymmorphical operation of a glide along the c-axis.
Under these symmetrical operations $\hat{U}$, the Hamiltonian of the black phosphorus, H(k), satisfies:
\begin{equation}
    H(\bf{k}) = U^{*}(k)H(\hat{U}\bf{k})U(\bf{k}).
\end{equation}
At $k_z=\pi/c$, or the Brillouin zone surface containing the Z point (0, 0, $\pi/c$),
the energy band structures of a nonsymmorphical crystal may be connected, leading to the
degeneracy \cite{Lee2003Numerical}.  Such a connection and degeneracy
lead to the fourfold-degenerate Dirac points \cite{weng2014transition,gibson20143d}.
\\

\begin{table*}[!t]
\caption{Extremal cross-section area $S_{F}$ of electron and hole pockets at pressure 2.0 GPa.
$\beta$, $\alpha$ and $\alpha^*$ stand for a Dirac dispersion hole pocket at $Z$-$\Gamma$ line,
a normal electron pocket and a Dirac dispersion electron pocket at $Z$-$M$ line, respectively.}
\begin{threeparttable}
\begin{tabular}{cccccccc}
\hline\hline
    Pockets type         & $\beta$ (hole)  \hspace{1cm}     & $\alpha$ (electron)    \hspace{1cm}   &$\alpha^{*}$ (electron)                  & \\
\hline
                         & $1.06$\tnote{a}   \hspace{1cm}    & $0.4$\tnote{a}     \hspace{1cm}     & $0.23$\tnote{a}                    \\
$S_{F}$($nm^{-2}$)       & $(0.42)$\tnote{b}  \hspace{1cm}    & $(0.22)$\tnote{b}   \hspace{1cm}    & -                        \\
                         & -                 \hspace{1cm}    &  -                  \hspace{1cm}    & $(0.16)$\tnote{c}     \\
\hline
Inequivalent numbers    & 2          \hspace{1cm}     &  4             \hspace{1cm}            & 2                         \\
\hline
Total $S_{F}$($nm^{-2}$)     & 2.12        \hspace{1cm}    & 1.60          \hspace{1cm}             & 0.46                       \\
\hline\hline
\end{tabular}
     \begin{tablenotes}
        \footnotesize
        \item[a] Ours,     the plane (0, $k_{y}$, $k_{z}$)    \\
        \item[b] Ref. [7], the plane ($k_{x}$, $k_{y}$, 0), with the magnetic field $H$=20 T parallelled to the $c$-axis and under the pressure of 2.0 GPa.   \\
        \item[c] Ref. [7], the plane ($k_{x}^{'}$, $k_{y}$, $k_{z}$), where the direction of $k_{x}^{'}$ is $18^{\circ}$ away from the $k_{x}$  direction,
        with $H$=20 T and under the pressure of 2.4 GPa.   \\

     \end{tablenotes}
\end{threeparttable}

\end{table*}

\begin{table*}[!t]
\caption{Effective masses of black phosphorus under various hydrostatic pressures
and comparison with Ref.\cite{Qiao2014High,Morita1989Electron}.}
\begin{tabular}{cccccccc}
\hline\hline
    & P(GPa) & $m^*_{x}/m_{0}$ & $m^*_{y}/m_{0}$ & $m^*_{z}/m_{0}$  \\
\hline
  & $0.00$ & $0.14(0.12[11], 0.08[37])$ & $1.26(1.15[11], 1.03[37])$ &
   $0.16(0.15[11], 0.13[37])$   \\
e & $0.50$ & $0.09$ & $1.24$ & $0.14$  \\
  & $1.00$ & $0.05$ & $1.21$ & $0.13$ \\
\hline
  & $0.00$ & $0.12(0.11[11], 0.07[37])$ & $0.90 (0.71[11], 0.65[37]) $ &
   $0.34 (0.30[11], 0.28[37])$   \\
h & $0.50$ & $0.08$ & $0.82$ & $0.32$  \\
  &$1.00$  & $0.04$ & $0.78$ & $0.31$ \\
\hline\hline
\end{tabular}
\end{table*}

%


\section{V. Fermi surfaces  evolution with pressures:}
\begin{figure}[!tb]
\noindent \begin{centering}
\includegraphics[width=6.0cm]{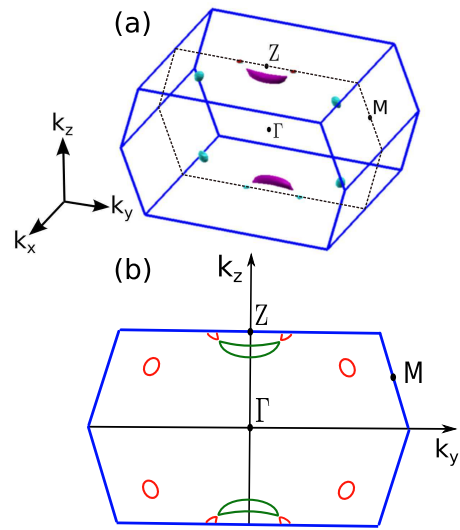}
\par\end{centering}
\caption{(Color Online) Fermi surface (a) and its projection (b) of bulk black phosphorus under the
pressure of 2.0 GPa.  The projection on the plane (0, $k_{y}$, $k_{z}$) is shown.  The two pockets
in the $\Gamma$-$Z$ line are hole type, and the others are electron type.}
\label{Fig8-projection}
\end{figure}

From the preceding studies we have shown that critical pressure $P_{c}$ is a {\it Lifshitz} point, which
the topology of Fermi surface of bulk black phosphorus changes crucially.
The Fermi surfaces and its cross-section of black phosphorus under the pressure
of 2.0 GPa are displayed in Fig. \ref{Fig8-projection}. There are
three different kinds of Fermi pockets: two hole pockets arising from
the upward shift of two Dirac cones in the $\Gamma$-$Z$ line above E$_F$,
and two tiny electron pockets on the upper and lower surface of the Brillouin zone
arising from the sink of two Dirac cones in the $M$-$Z$ line below E$_F$; another
four electron-type pockets in the $Z$-$M$ line come from the sink of the bands
below E$_F$. These four Fermi pockets do not originate from the Dirac cones.

The cross-section Fermi surface on the plane (0, $k_{y}$, $k_{z}$) at 2.0 GPa is displayed
in Fig. \ref{Fig8-projection}(b). This shows that the Dirac points and the centers of the Fermi pockets are
on the $k$$_x$=0 plane.  The extremal cross-section area $S_{F}$ of the electron and hole pockets
in the plane (0, $k_{y}$, $k_{z}$) are estimated.
Our results show  $S_{F}$ are 0.40, 0.23 and 1.06 $nm^{-2}$ for a large electron pocket ($\alpha$),
a tiny electron pocket ($\alpha^{*}$) and a hole pocket ($\beta$), respectively, listed in Table I.
As for Shubnikov-De Haas oscillation experiment \cite{xiang2015pressure}, it reports that 0.22
for $\alpha$ and  0.42  $nm^{-2}$ for $\beta$ in the plane
($k_{x}$, $k_{y}$, 0), with the magnetic field $H=20$ T parallelled to the $c$-axis and under the pressure of 2.0 GPa;
while $S_{F}$ for $\alpha^{*}$ is 0.16 $nm^{-2}$ in  ($k_{x}^{'}$, $k_{y}$, $k_{z}$) with $H=20$ T
and under the pressure of 2.4 GPa.
Our theoretical results have the same order magnitude with the experiment \cite{xiang2015pressure}.
With consideration of the number of two types of pockets (2 for $\beta$, 4 for  $\alpha$ and 2 for  $\alpha^*$ )
in the first Brillouin zone, the total $S_{F}$s of electron pockets and hole pockets are approximately equal to
2.00 $nm^{-2}$. The numbers of two kinds of carriers keep identical, indicating a complete compensation.
These theoretical data show that the system is semimetal state, in good agreement with the results by the
field dependence of transverse magnetoresistance experiment \cite*{xiang2015pressure}.
Also, the Dirac fermions and normal fermions coexist in the black phosphorus under the
pressure larger than $P_{c}$, providing a complete novel quantum system.

\begin{figure}[!tb]
\noindent \begin{centering}
\includegraphics[width=8.0cm]{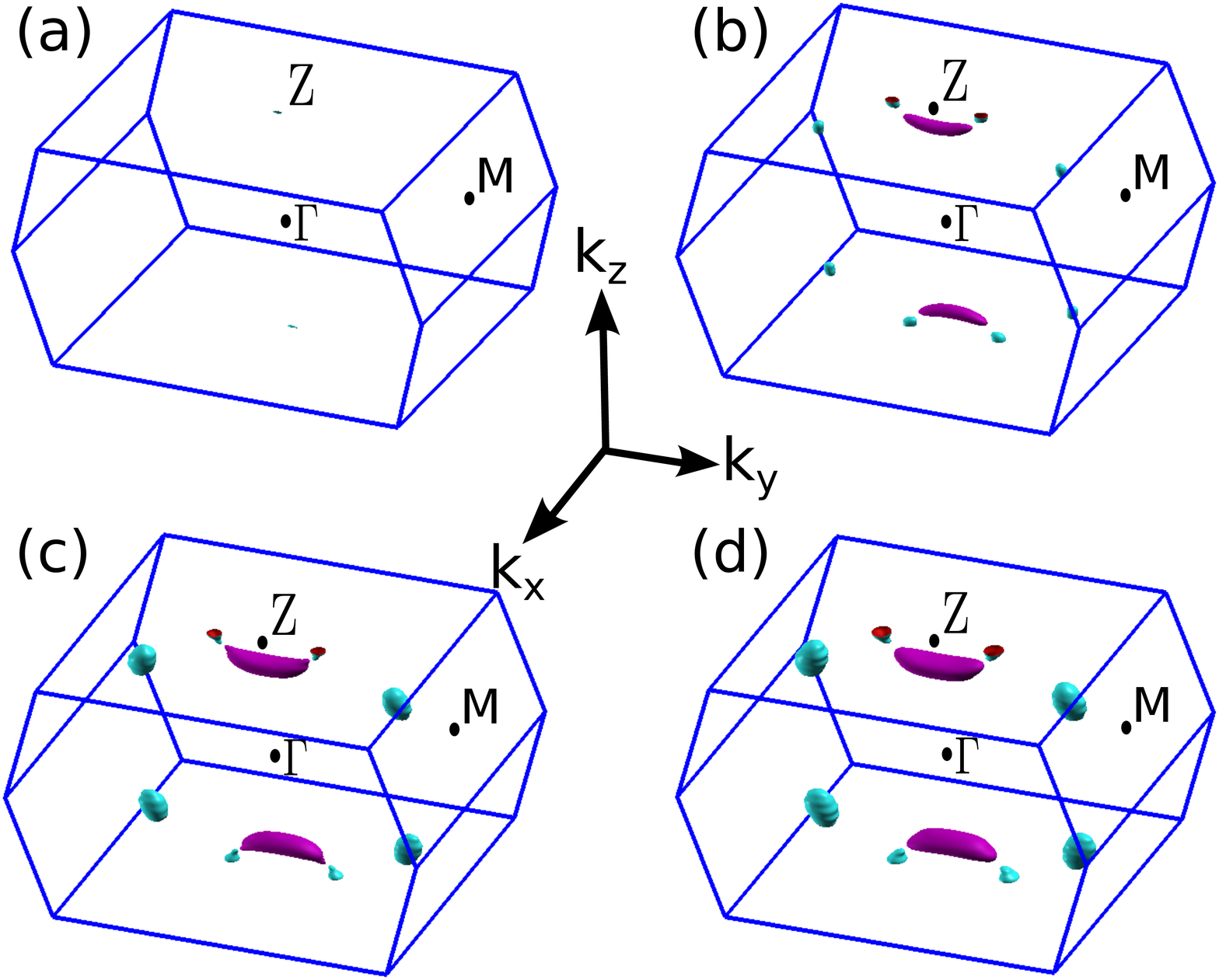}
\par\end{centering}
\caption{Calculated Fermi surfaces for  bulk black  phosphorus from 1.23 GPa (a), 1.5 GPa (b),
3.0 GPa (c) and 4.0 GPa (d).}
\label{Fig9-FS}
\end{figure}

The pressures can modify the electronic structures, leading to the interesting evolution of Fermi surfaces,
and driving significant change in the topology of electronic structures, {\it i. e.}, the {\it Lifshitz} transition .
We present Fermi surfaces topology of bulk black phosphorus under different pressures in Fig. \ref{Fig9-FS}.
It is shown that Fermi surface is null before the critical pressure for the semiconductor-semimetal transition.
With increasing the pressure to 1.5 and 3.0 GPa, up to 4.0 GPa, the volumes of hole Fermi pockets in
$\Gamma$-$Z$ line gradually expands, and electron Fermi pockets in $Z$-$M$ line grow largely,
in consistent with the resent experimental results \cite{xiang2015pressure}.
The volumes of electron and hole pockets become large with increasing pressure, indicating that more and
more hole and electron carriers are produced in the system.
\\

\section{VI. Anisotropic Physical Properties}

The anisotropic electronic structures in black phosphorus result in highly orientation-dependent
properties, such as the effective masses of carriers and Fermi velocities, {\it etc.} It displays the
hydrostatic pressure dependence of the Fermi
velocities ($V_{F}$) of the hole and electron carriers by the PBE potential shown in Fig. \ref{Fig10-VF}(a)
and the mBJ one in Fig. \ref{Fig10-VF}(b), respectively.
We  focus on the Fermi velocity of Dirac cones near the Fermi level, which is important for
the fundamental properties of the system. Note that we calculate the average  Fermi velocity from
two branches of Dirac cone. It has shown that the considerable pressure dependence of the Fermi
velocity both for the PBE and the mBJ results, while, the pressure dependence of V$_F$ by
the mBJ potential is significantly larger than that by the PBE potential.
Of course, more elaborated experiments are expected to confirm our theoretical details in the future.

In addition, It is seen that the Fermi velocities
monotonically increase with increasing the hydrostatic pressure.  The velocities along the
$Z$-$\Gamma$ direction are almost twice of those along the $Z$-$M$ direction, showing
strong transport anisotropy. Meanwhile, the effective masses of the hole and electron carriers
also display considerable anisotropy.
As listed in Table II, with increasing hydrostatic pressure, the effective masses of electron
carriers in the y- and z-axis almost do not alter, and the effective masses of hole
carriers along the z-axis change slightly. However, the effective masses along the x-axis
decrease to about one-third of the ambient pressure, demonstrating strong pressure modulation.

\begin{figure}[!htbp]
\noindent \begin{centering}
\includegraphics[width=4.0cm]{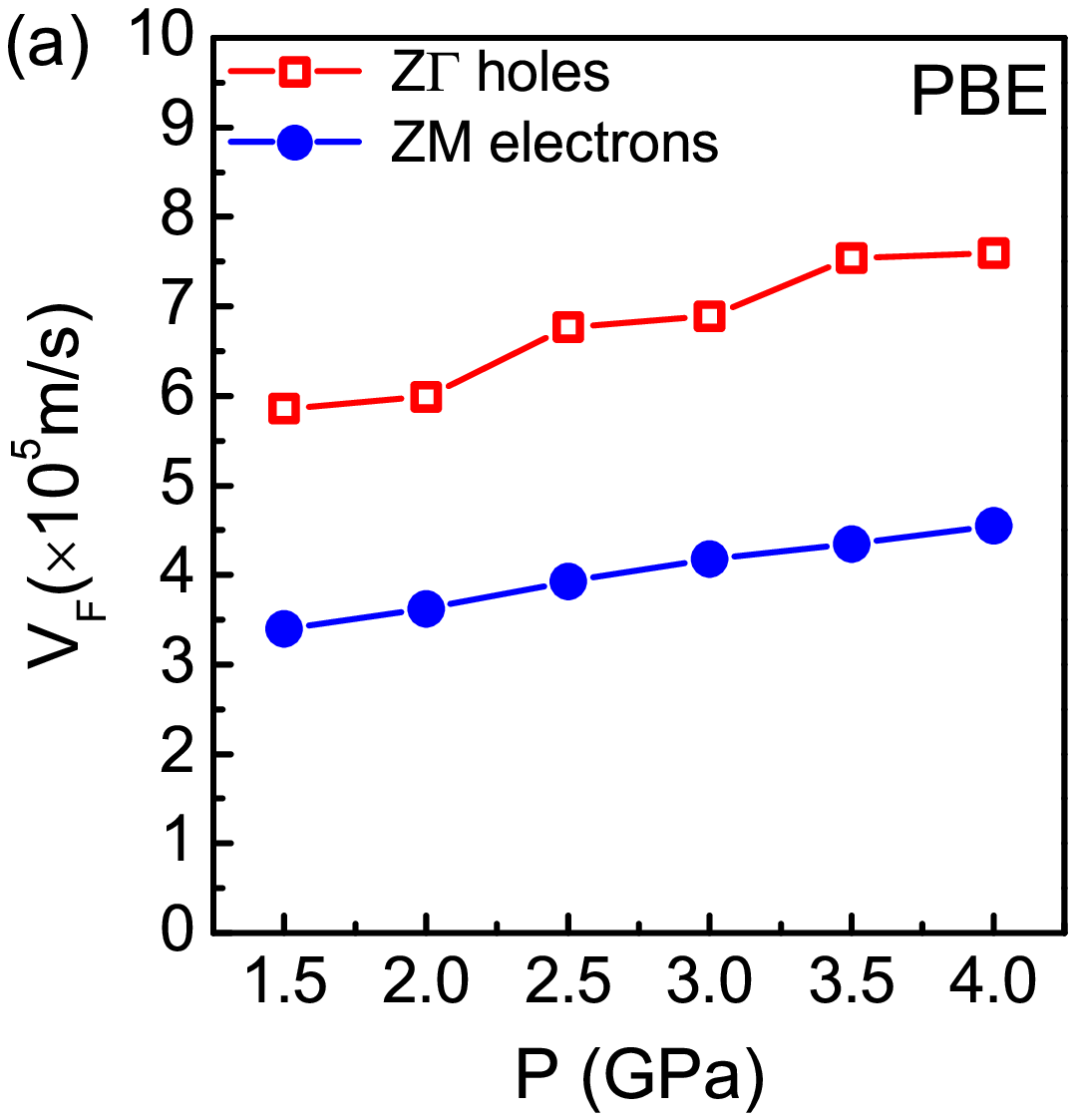}
\includegraphics[width=4.0cm]{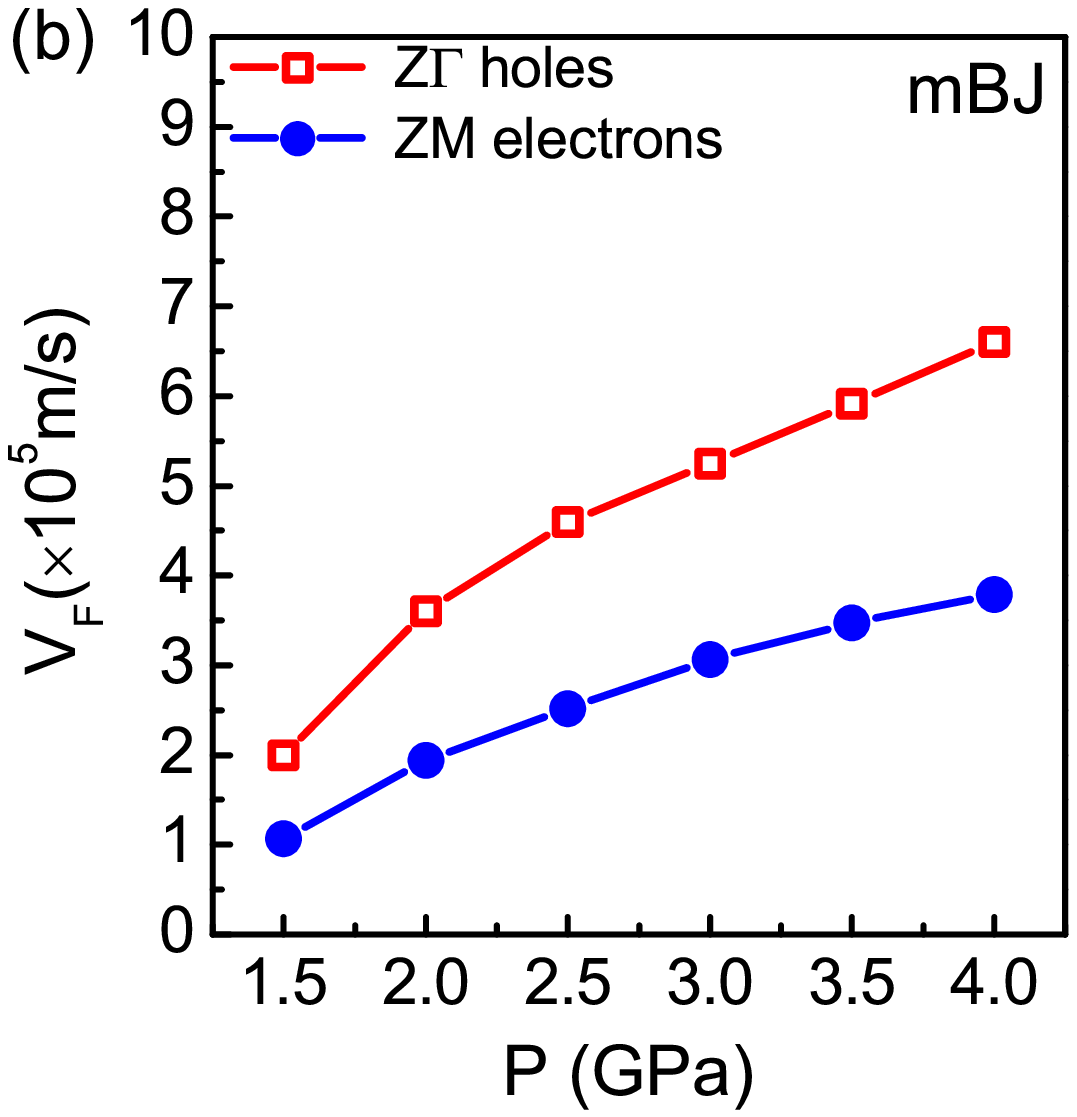}
\par\end{centering}
\caption{ (Color Online) Dependence of the average Fermi velocity of  bulk black
phosphorus under the pressures from 1.5 GPa up to 4.0 GPa by the PBE (a) and the mBJ potentials (b),
respectively. }
\label{Fig10-VF}
\end{figure}

\section{VII. Summary}
This work presents the nontrivial influence of hydrostatic pressure on bulk black phosphorus, which is distinct
different from the strain-induced effect \cite{rodin2014strain}. The hydrostatic presure shortens
the distance of two P atomic layers and  the height of P atomic layers, which
results in considerable changes in electronic structures and Fermi surface topology,
hence a {\it Lifshitz} transition.
It also leads to the appearance of 4 twofold-degenerate Dirac cones at a critical pressure $P$$_{c}$,
implying that black phosphorus under hydrostatic pressure becomes a 3D Dirac semimetal.
More recent work suggested that the these Dirac points may form unusual nodal ring in pressured
black phosphorus \cite{Weng2015}
These show that pressure controlled layered black phosphorus may
have potential application in optoelectronic and electronic device.
\\

\begin{acknowledgements}

The authors L.-J. Zou and D.-Y. Liu thank X.-G. Gong for his hospitality during their visit to the
Department of  Physics, Fudan University.  This work was supported by the NSF of China under
Grant No. 11474287, 11274310, 11104274 and the key project 11534010.
Numerical calculations were performed at the Center for Computational Science of CASHIPS.
\end{acknowledgements}


\renewcommand\thefigure{A\arabic{figure}}

\section{VIII. Appendix}

 To make sure that pressured structures are thermodynamically stable, we calculate the phonon spectra 
of bulk black phosphorus for our theoretical structures under the hydrostatic pressures
from 0 GPa to 4 GPa, as shown in Fig. \ref{FigA1-phonon}. Our results show that the 
phonon spectra of these structures have no any imaginary frequency, implying that the optimized geometries really locate at the minimum point of the potential surface.

\begin{figure}[!t]
\includegraphics[width=8.5cm]{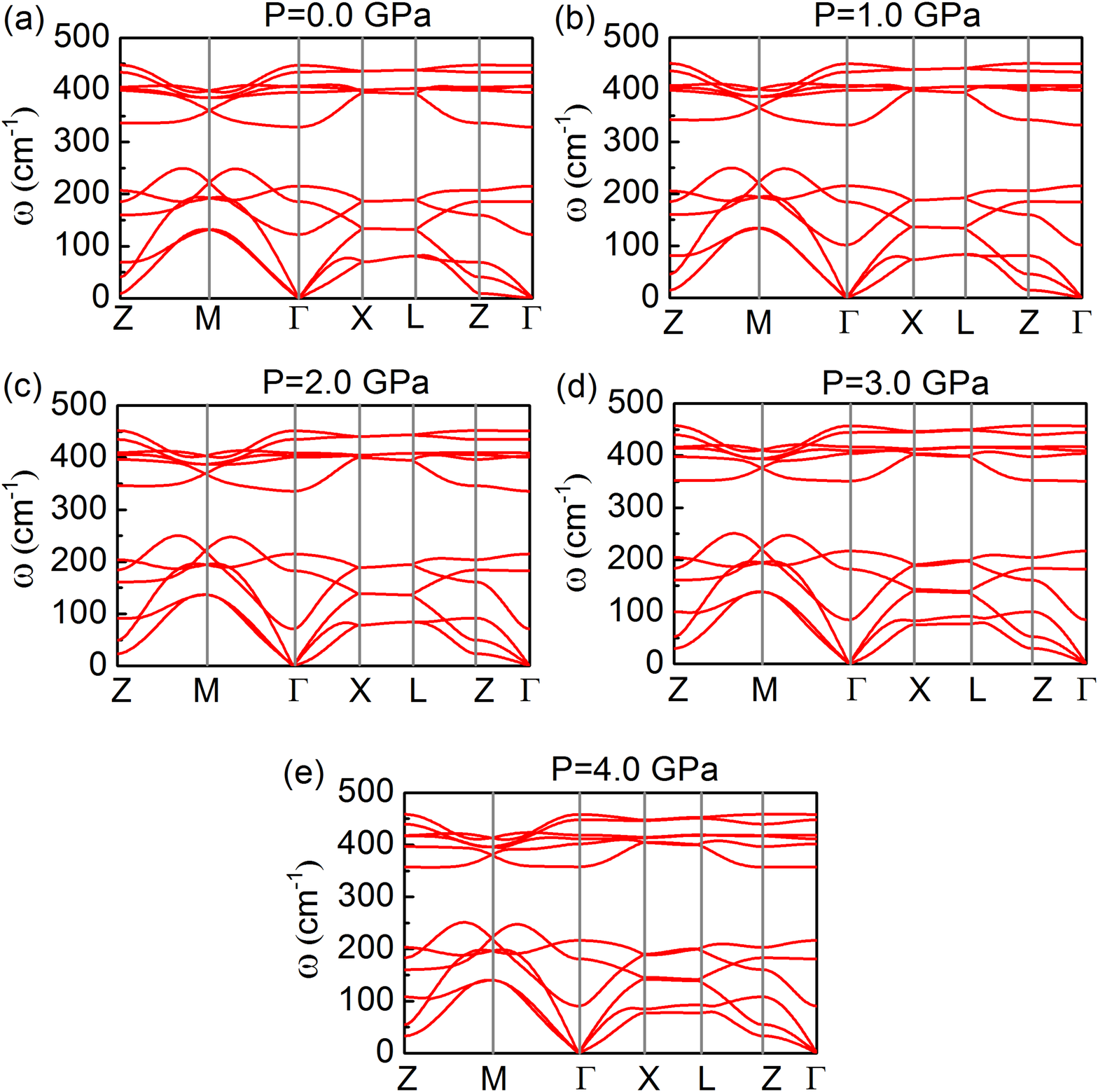}
\setcounter{figure}{0}
\caption{Calculated phonon dispersion of  bulk black  phosphorus  under pressures of $P$=0, 1, 2, 3, 
and 4 GPa.}
\label{FigA1-phonon}
\end{figure}



\end{document}